\documentclass[11pt]{article}  
\usepackage[utf8]{inputenc}     
\usepackage[T1]{fontenc}        
\usepackage{lmodern}            
\usepackage{amsmath, amssymb}   
\usepackage[margin=1in]{geometry}  
\usepackage{graphicx}           
\usepackage{hyperref}           %
\usepackage{authblk}  
\newcommand{\onlinecite}[1]{\cite{#1}}

\title{A model for Reynolds-dependent linear instability in wall-bounded shear flows}

\author[1,2]{John O. Dabiri}
\author[1]{Anthony Leonard}

\affil[1]{Graduate Aerospace Laboratories, California Institute of Technology, Pasadena, CA 91125}
\affil[2]{Department of Mechanical and Civil Engineering, California Institute of Technology, Pasadena, CA 91125}
\date{}

\begin{document}
\maketitle

\begin{abstract}
Recent work demonstrated that alternative models to the ``no-slip'' boundary condition for incipient flow perturbations can produce linear instabilities that do not arise in the classical formulation. The present study introduces a Robin-type boundary condition with explicit Reynolds-number dependence, which leads to a more physically realistic transition to instability in wall-bounded shear flows. The results show that instability occurs within a Reynolds number range consistent with empirical observations. A physical interpretation of the modified boundary condition in terms of near-wall vorticity dynamics is discussed.
\end{abstract}

Recent work \cite{Dabiri2024} demonstrated that alternative models to the ``no-slip'' boundary condition for incipient flow perturbations can produce linear instabilities that do not arise in the classical formulation. One of the models in that work (Perturbation Model II in Ref.~\onlinecite{Dabiri2024}) exhibited a transition to linear instability in a range of Reynolds numbers that is consistent with empirical observations of turbulence transition. However, that model only predicted instability in a relatively narrow range of parameter space, and its derivation relied upon an \emph{a priori} assumption of nearly neutral stability for all flow perturbations.  

In this note, we derive a more straightforward, Robin-type boundary condition for the near-wall behavior of incipient flow perturbations. While the model shares the form of Perturbation Model I in Ref.~\onlinecite{Dabiri2024}, a Reynolds-number-dependent transition to linear instability is achieved via incorporation of the critical layer thickness scaling in the model. 

Similar to Perturbation Model I in Ref.~\onlinecite{Dabiri2024}, we consider an alternative to the no-slip boundary condition of the form

\begin{equation} \label{eq:Redependentslip}
    \tilde{u}(\mathbf{x_{wall}},t) \mp S\, \tilde{u}'(\mathbf{x_{wall}},t) = 0
\end{equation}

\noindent where $\tilde{u}$ is the streamwise component of the velocity perturbation amplitude, $\tilde{u}' = \partial \tilde{u}/\partial y$ for plane Couette and Poiseuille flows, $\tilde{u}' = \partial \tilde{u}/\partial r$ for Hagen-Poiseuille pipe flow, $S$ is a characteristic length scale of the wall shear (normalized by the channel half-width or pipe radius), and the sign $\mp$ applies to the location of the wall $\mathbf{x_{wall}}$ at $y$ (or $r$) $= 1$ and $y = -1$, respectively.

Physically, this boundary condition places no specific, quantitative constraint on the magnitude of flow perturbations along the wall. Instead, it only requires that any non-zero perturbation velocity along the wall is in the same direction---and proportional to---the corresponding perturbation shear exerted on the fluid by the wall. Whereas the constant of proportionality $S$ was treated as a free parameter in Perturbation Model I, here we instead derive the expected scaling of this parameter with Reynolds number. We begin with the Orr-Sommerfeld equation for incipient perturbations:

\begin{equation}\label{eq:OSequation}
\Bigl[(-i\omega + i\alpha U)(\mathcal{D}^2 - k^2)-i\alpha U''-\frac{1}{Re}(\mathcal{D}^2 - k^2)^2\Bigr]\tilde{v} = 0
\end{equation}

\noindent where $\alpha$ and $\beta$ are the streamwise and spanwise wavenumbers, respectively, $k^2 = \alpha^2 + \beta^2$, and $Re = U_0L_0/\nu_0$ is the Reynolds number. Here, $U_0$ is taken as the maximum velocity of the base flow in the domain, and $L_0$ is the half-width of distance between the solid walls of the Couette or Poiseuille flow. The operator $\mathcal{D}$ and the prime both denote a derivative with respect to the wall-normal $y-$coordinate direction.

Taking $S$ as the characteristic wall-normal length scale of the perturbation shear, we seek to balance the dominant viscous and inertial terms in the vicinity of the critical layer, i.e., where the perturbation phase speed $\mathbb{R}(c)$ is equal to the local base flow speed. The highest-order viscous term is

\begin{equation}\label{eq:viscousscaling}
\frac{1}{Re}(D^2-k^2)^2\tilde{v} \sim \frac{1}{Re}\left(\frac{1}{S^2}\right)^2 \tilde{v} = \frac{1}{Re}\frac{1}{S^4}\tilde{v} .
\end{equation}

\noindent where the $k^2$ term has been neglected in comparison with the spatial derivatives, which are assumed to be large near the wall.

Similarly, the inertial term scales as

\begin{equation}\label{eq:inerialscaling}
(-i\omega + i\alpha U)\mathcal{D}^2\tilde{v} \sim i\alpha\,S\,\frac{1}{S^2}\tilde{v} \sim \frac{1}{S}\tilde{v}
\end{equation}

\noindent assuming the streamwise wavenumber $\alpha \sim \mathcal{O}(1)$ and the velocity contrast $U-\mathbb{R}(c) \sim S$ in a linear shear assumption. Balancing the viscous and inertial terms in equations (\ref{eq:viscousscaling}) and (\ref{eq:inerialscaling}) gives

\begin{equation}\label{eq:balance}
\frac{1}{Re}\frac{1}{S^4}\tilde{v} \sim \frac{1}{S}\tilde{v}.
\end{equation}

\noindent or, equivalently,

\begin{equation}\label{eq:Sscaling}
S \sim Re^{-1/3}
\end{equation}

The scaling in equation (\ref{eq:Sscaling}) predicts that as the Reynolds number increases, influence of perturbation wall shear is confined closer to the wall. It should be noted that, in general, critical layers need not be located near the wall. However, the present model assumes that the flow perturbation arises at the fluid-solid interface, hence it is confined to that region of the flow. We revisit this assumption via examination of perturbation phase speed $\mathbb{R}(c)$ shortly.  Based on the preceding derivation, the modified boundary condition for incipient perturbations is given as 

\begin{equation} \label{eq:Redependentslip2}
    \tilde{u}(\mathbf{x_{wall}},t) \mp \frac{C\,\tilde{u}'(\mathbf{x_{wall}},t)}{Re^{1/3}} = 0
\end{equation}

\noindent where $C$ is a dimensionless wall layer scaling parameter of $\mathcal{O}(1)$.

\begin{figure*}
    \centering
    \includegraphics[width=0.55\textwidth]{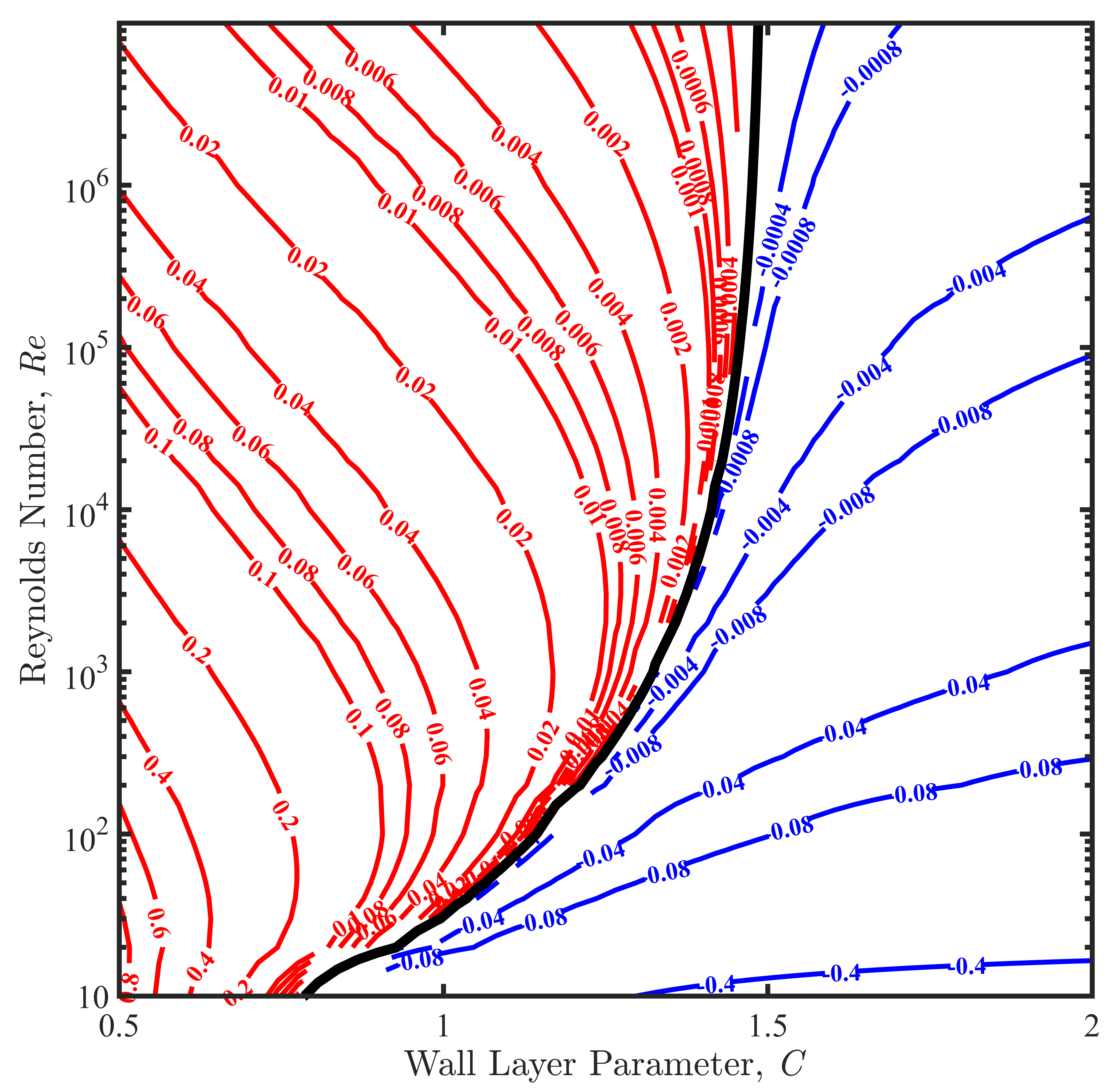}
    \caption{Contour map of the maximum Orr-Sommerfeld eigenvalue imaginary part $\mathbb{I}[\hat{\omega}]$ versus wall layer scaling parameter $C$ and Reynolds number $Re$ for plane Couette flow. Blue contours indicate regions of linear stability, and red contours indicate regions of linear instability. Black contours indicate neutral stability boundaries. Wavenumber is $(\alpha, \beta) = (1, 0)$.}
    \label{fig:map}
\end{figure*}

The Orr-Sommerfeld equation was solved using this modified boundary condition for plane Couette flow, plane Poiseuille flow, and Hagen-Poiseuille pipe flow using Chebyshev collocation with at least $N = 160$ modes (see details of the implementation in Ref.~\onlinecite{Dabiri2024}). The solutions for Couette flow were further verified semi-analytically by leveraging the fact that the flow also satisfies an Airy equation. Figure~\ref{fig:map} plots contours of the maximum eigenvalue imaginary part, i.e., $\mathbb{I}[\hat{\omega}]$, for Orr-Sommerfeld solutions corresponding to plane Couette flow using the perturbation boundary condition in equation (\ref{eq:Redependentslip2}) along with a no-penetration condition (i.e., $\tilde{v}(\mathbf{x_{wall}},t) = 0$). Values of the wall layer scaling parameter $C \sim \mathcal{O}(1)$ are shown on the abscissa. Within this range, the model predicts the onset of linear instability in a range of Reynolds numbers that is consistent with empirical observations. 

\begin{figure*}
    \centering
    \includegraphics[width=0.55\textwidth]{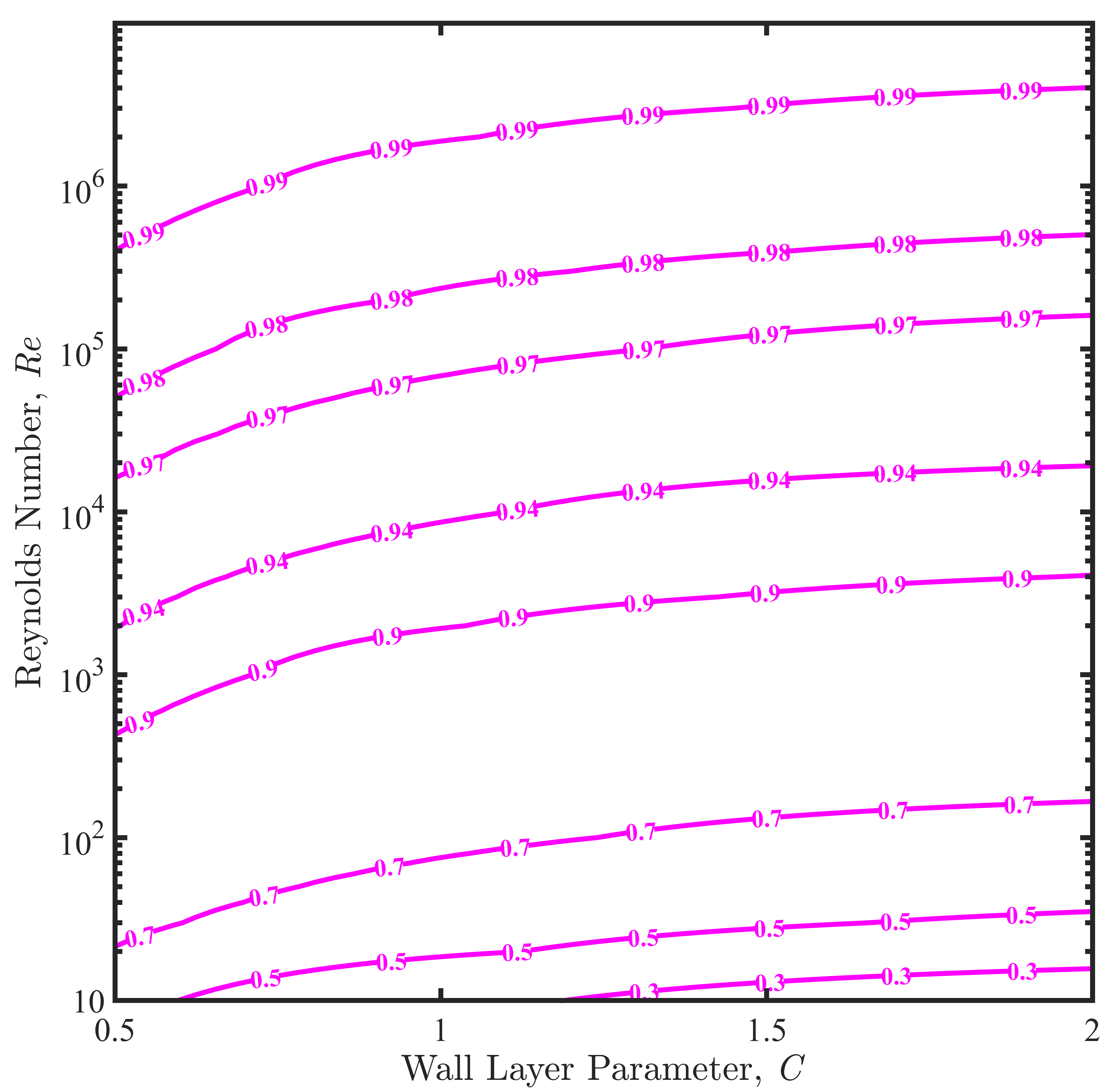}
    \caption{Contour map of the phase speed of the eigenmode with maximum Orr-Sommerfeld eigenvalue imaginary part $\mathbb{I}[\hat{\omega}]$ versus wall layer scaling parameter $C$ and Reynolds number $Re$ for plane Couette flow. Wavenumber is $(\alpha, \beta) = (1, 0)$.}
    \label{fig:phase}
\end{figure*}

An important difference between the predictions of this model and Perturbation Model I in Ref.~\onlinecite{Dabiri2024} is that the magnitude of the instabilities here is more modest and arguably more physically realistic (cf. $\mathbb{I}[\hat{\omega}] > 10^6$ in some cases of Perturbation Model I). Furthermore, examination of the phase speed of the instabilities confirms that as the Reynolds number increases, the phase speed of the disturbances approaches the wall speed (figure~\ref{fig:phase}). This result is consistent with the assumption in the boundary condition (i.e., equation \ref{eq:Redependentslip2}) that the critical layer is increasingly confined to the near-wall region at high Reynolds numbers.

\begin{figure*}
    \centering
    \includegraphics[width=0.475\linewidth]{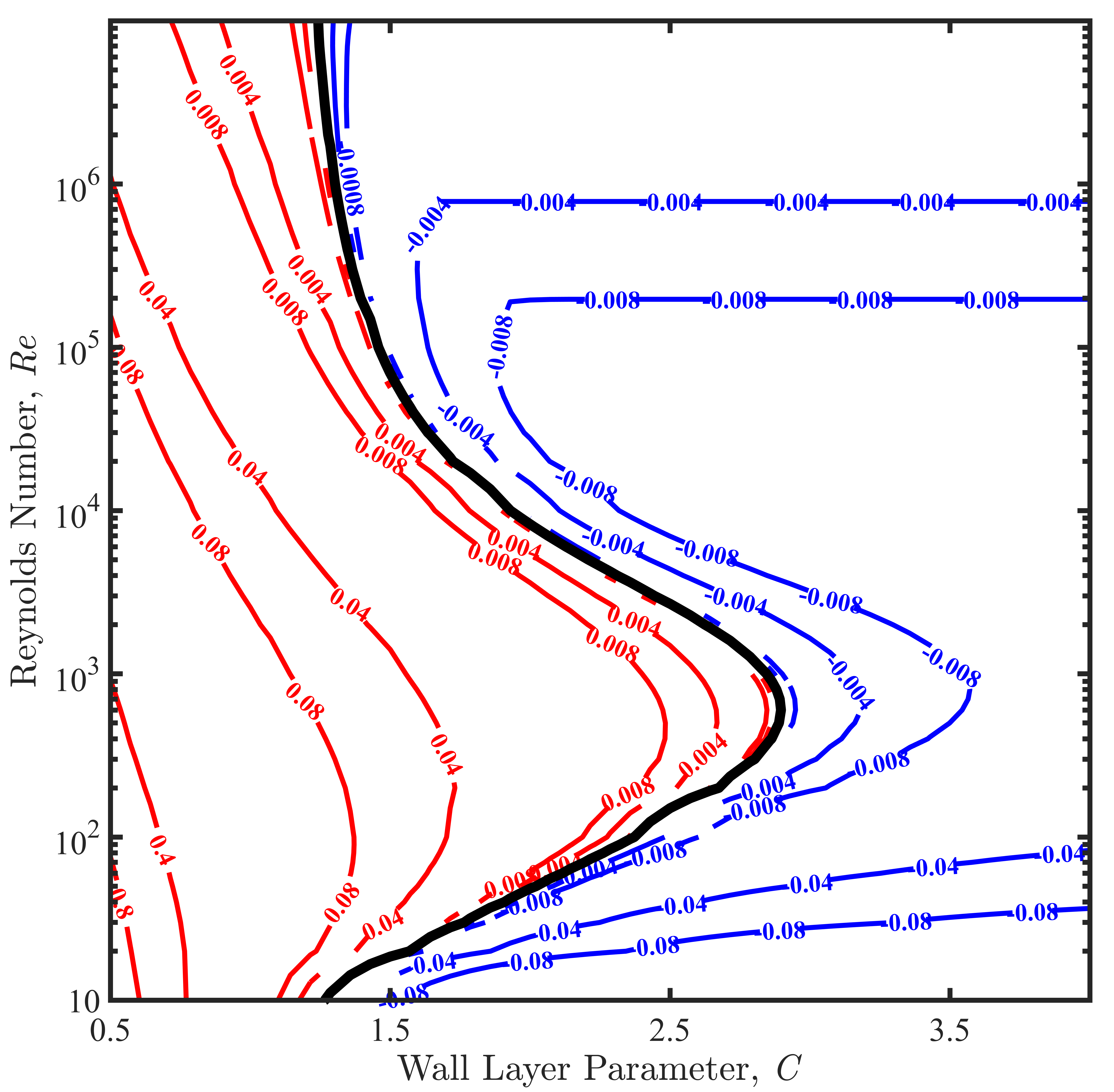}
    \includegraphics[width=0.475\linewidth]{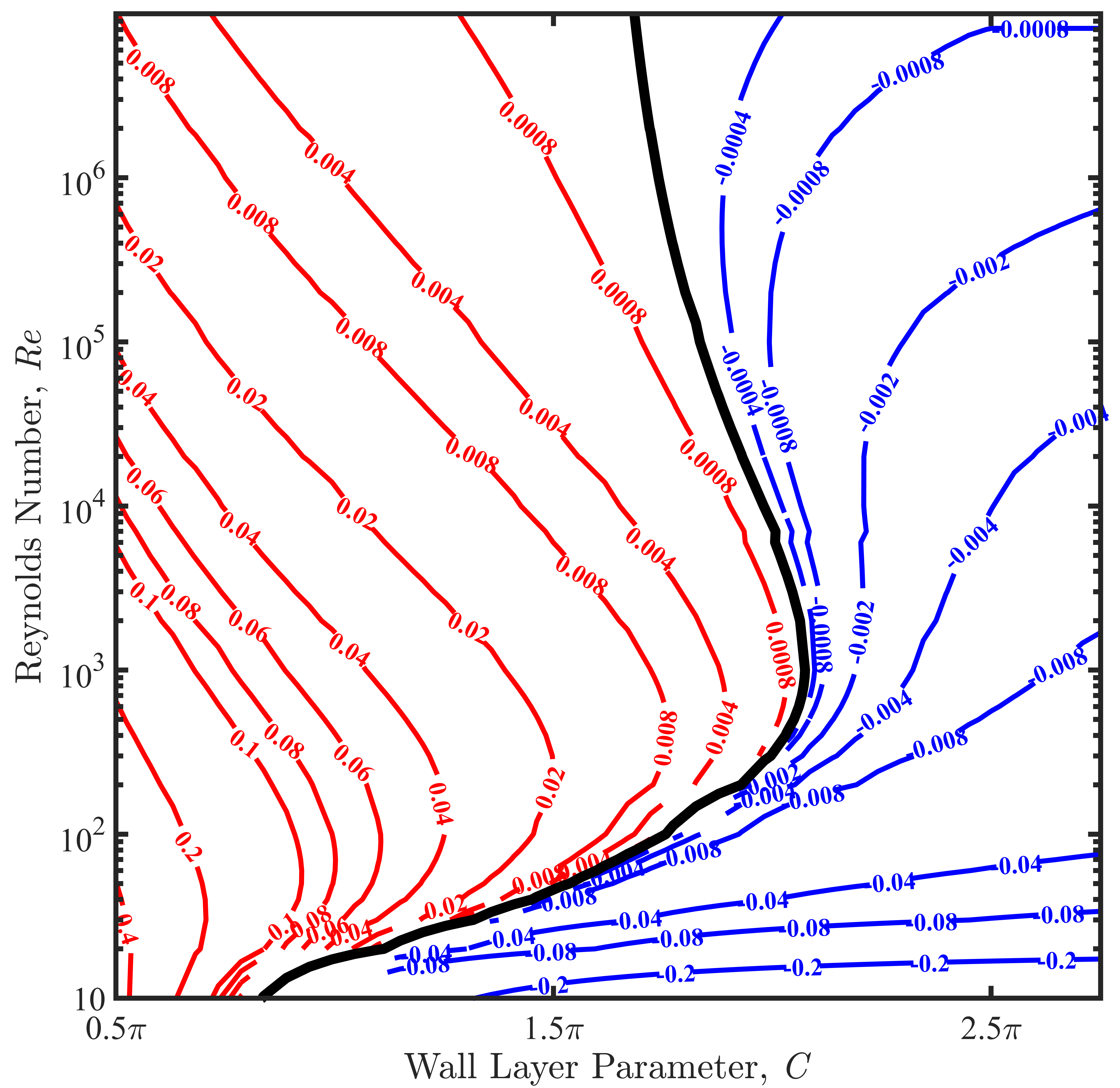}
    \caption{Contour maps of the maximum Orr-Sommerfeld eigenvalue imaginary part $\mathbb{I}[\hat{\omega}]$ versus wall layer scaling parameter $C$ and Reynolds number $Re$ for plane Poiseuille flow with $(\alpha, \beta) = (1, 0)$ (\emph{left}) and for Hagen-Poiseuille pipe flow with $(\alpha, n) = (1, 1)$ (\emph{right}).} 
    \label{fig:map_Re_C_planepoise}
\end{figure*}

\begin{figure*}
    \centering
    \includegraphics[width=0.475\linewidth]{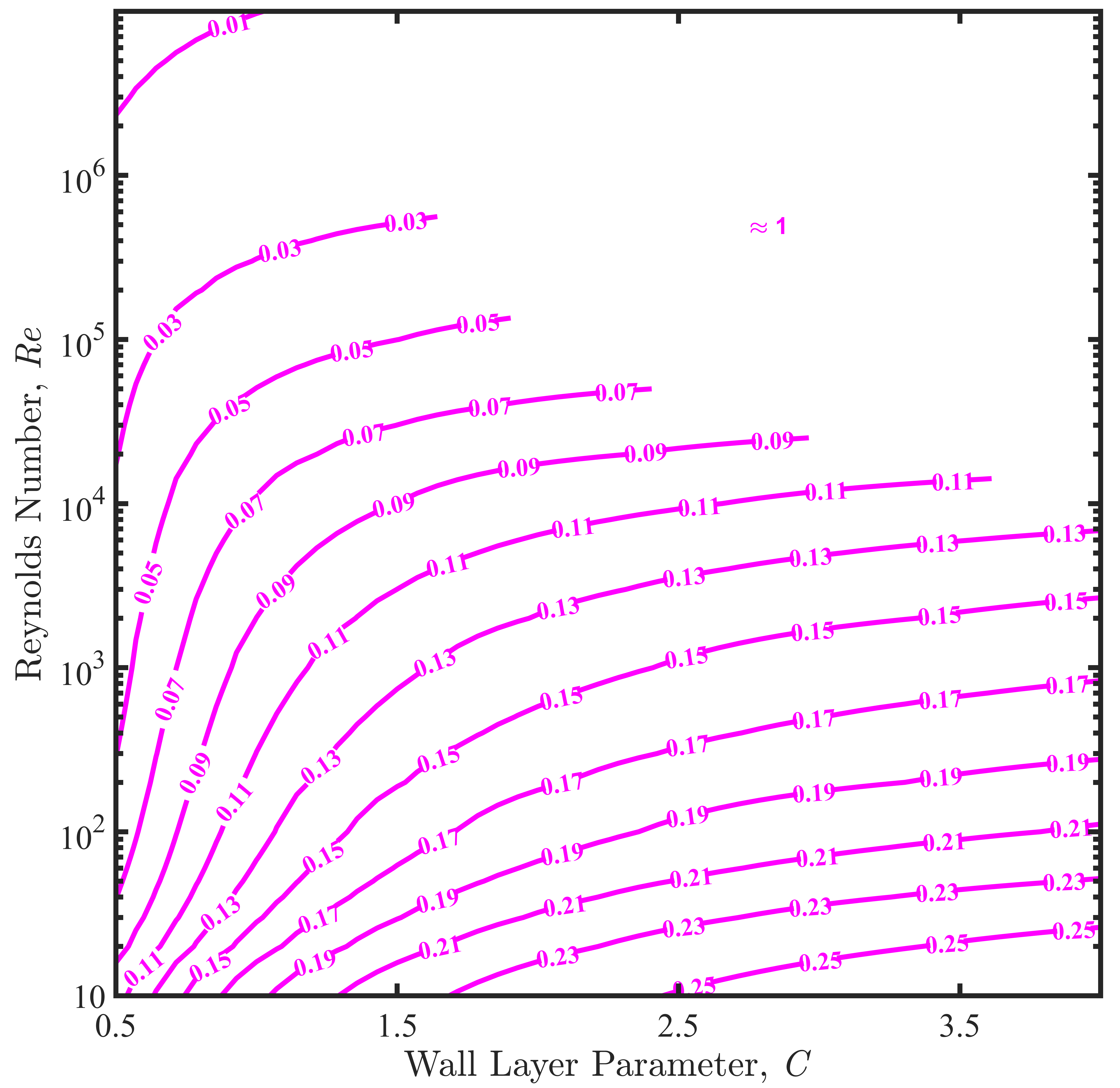}
    \includegraphics[width=0.475\linewidth]{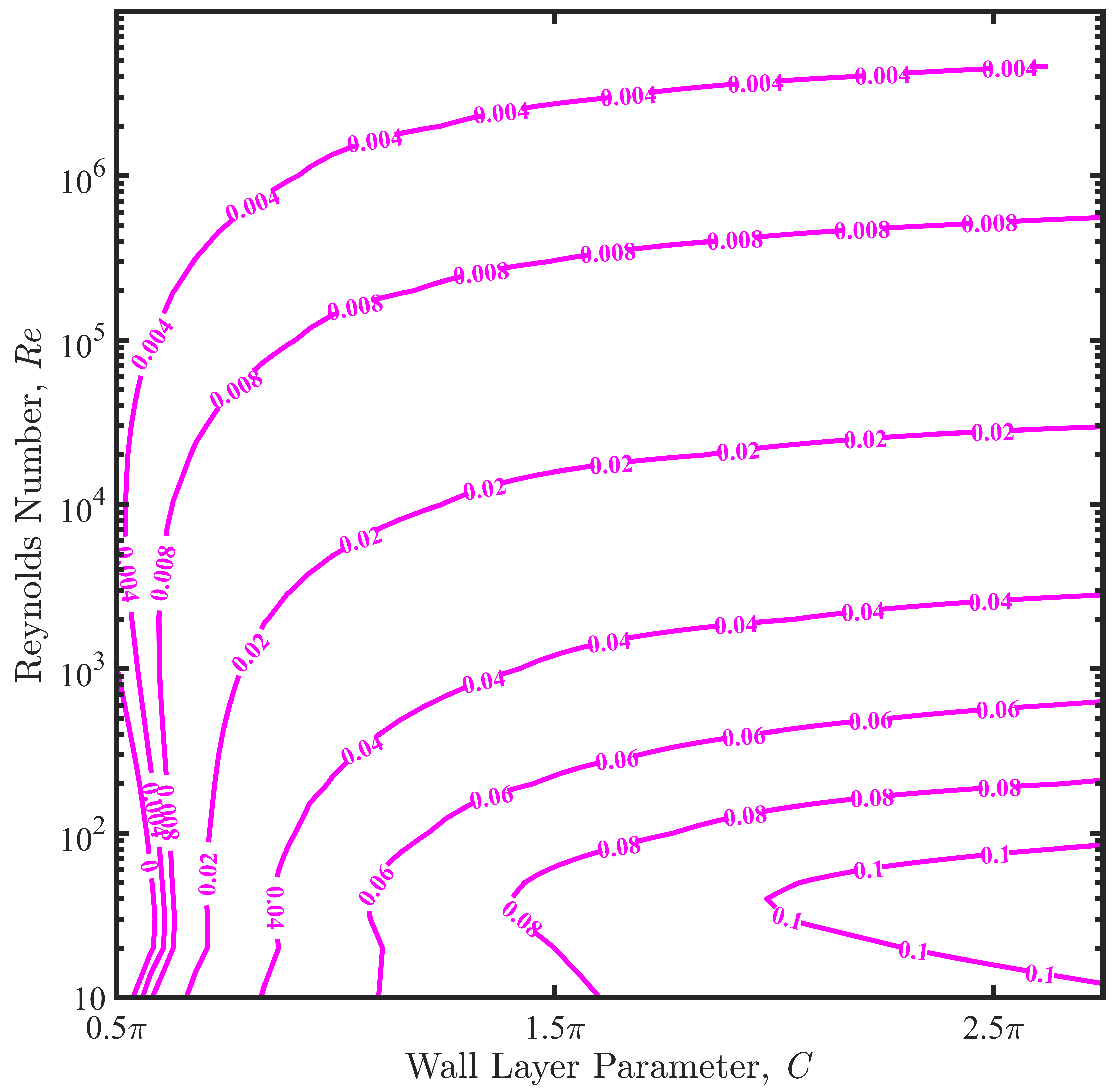}
    \caption{Contour maps of the phase speed of the eigenmode with maximum Orr-Sommerfeld eigenvalue imaginary part $\mathbb{I}[\hat{\omega}]$ versus wall layer scaling parameter $C$ and Reynolds number $Re$ for plane Poiseuille flow with $(\alpha, \beta) = (1, 0)$ (\emph{left}) and for Hagen-Poiseuille pipe flow with $(\alpha, n) = (1, 1)$ (\emph{right}).} 
    \label{fig:phase_Re_C_planepoise}
\end{figure*}

Similar results were observed in analysis of plane Poiseuille flow as well as for Hagen-Poiseuille pipe flow. As shown in figure \ref{fig:map_Re_C_planepoise}, the region of linear instability has a similar shape for both Poiseuille flows, particularly if the wall layer length scale parameter $C$ is scaled by a geometric factor $\pi$ in the pipe flow. As in the case of plane Couette flow, the phase speed of the disturbances approaches the wall speed as the Reynolds number increases (figure \ref{fig:phase_Re_C_planepoise}). For the Poiseuille flows, the wall speed is zero as it is stationary. A notable exception in the phase speed behavior occurs for the planar Poiseuille flow at large values of $C$ and $Re$. In this region of parameter space,  the phase speed abruptly changes to values approaching 1, i.e., the centerline flow speed. This phenomenon suggests a shift from wall-dominated eigenmodes to modes associated with the bulk flow. However, this shift occurs in a region of parameter space that is generally far from the transition to linear instability. Therefore it is not inconsistent with the model assumption that the wall boundary condition governs the occurrence of linear instability.

For larger values of $C$, i.e., where perturbation wall slip occurs in greater proportion to the perturbation wall shear, the threshold Reynolds number of linear instability increases significantly. For sufficiently large $C$, the flow is predicted to remain stable almost indefinitely. To be sure, it remains to be seen if a practical method exists to enhance the sensitivity of wall slip to perturbation shear such that $C$ can be intentionally increased. Nonetheless, it is plausible that the observed delay in turbulence transition reported in the literature for some carefully constructed experimental wall-bounded flows (e.g., \cite{Pfenninger1961}) is associated with an increase in the wall layer parameter $C$.

The hypothesized perturbation wall slip inherent in the present model is necessarily associated with the concurrent generation of near-wall perturbation vorticity. Specifically, because $\tilde{v}(\mathbf{x_{wall}},t) = 0$ to satisfy the no-penetration condition, $\partial \tilde{v}(\mathbf{x_{wall}},t)/\partial x = 0$ by definition. The spanwise perturbation vorticity at the wall is therefore given as $\tilde{\omega}_z(\mathbf{x_{wall}},t) = -\tilde{u}'(\mathbf{x_{wall}},t)$, and the boundary condition equation (\ref{eq:Redependentslip2}) can be re-written as

\begin{equation} \label{eq:Redependentslip3}
    \tilde{\omega}_z(\mathbf{x_{wall}},t) = \mp \frac{Re^{1/3}}{C} \tilde{u}(\mathbf{x_{wall}},t)
\end{equation}

Hence, perturbation vorticity is generated at the wall in direct proportion with the perturbation wall slip. While the bulk perturbation vorticity transport equation is unchanged by the modified boundary condition, the perturbation streamfunction $\tilde{\psi}$ (where $\nabla^2 \tilde{\psi} = -\tilde{\omega}_z$) is no longer constrained to satisfy $\partial \tilde{\psi}/\partial y  = 0$ (or $\partial \tilde{\psi}/\partial r  = 0$)  at the boundary. This allows for the existence of perturbation vorticity field solutions that are unencumbered by the strict requirement that the wall-normal gradient of $\tilde{\psi}$ vanish at the wall.

The conventional no-slip condition also limits the persistence of any near-wall perturbation vorticity due to viscous diffusion and dissipation of the associated perturbation velocity gradient at the wall. By contrast, the existence of finite perturbation slip at the wall can enable perturbation vorticity to move along the wall and extract energy from the mean shear via vortex tilting and stretching. Previous studies of vorticity transport in turbulent flows have demonstrated the existence of conditions under which local vorticity perturbations can experience significant amplification by this mechanism \cite{Girimaji1990,Leonard2002}. Further exploration of the stability of spatially localized vorticity perturbations using similar methods can complement the present study of spatially non-local perturbations that are instead localized in wavenumber space.

\bibliographystyle{unsrt}
\bibliography{references}

\end{document}